%% file: main.tex
\documentclass[10pt,conference]{IEEEtran}
\IEEEoverridecommandlockouts

\usepackage{cite}
\usepackage{amsmath,amssymb,amsfonts}
\usepackage{textcomp}
\usepackage{url}
\usepackage[colorinlistoftodos,prependcaption,textsize=tiny]{todonotes}
\usepackage{xcolor}
\usepackage{enumitem}

\usepackage{subfigure}
\usepackage{booktabs}
\usepackage{balance}
\usepackage{hyperref}

\usepackage{graphicx}
\usepackage{algorithm}
\usepackage{algorithmic}
\usepackage{xspace}
\usepackage{multirow}
\usepackage{soul}

\usepackage{pifont}
\usepackage[most]{tcolorbox}

\newtcolorbox{mybox}{
  coltext=black,    
  boxrule=1.2pt,    
  arc=2pt,           
  width=0.495\textwidth, 
  center            
}

\newcommand{\etal}{{\em et al.}\xspace}
\newcommand{\ie}{{\em i.e.},\xspace}
\newcommand{\eg}{{\em e.g.},\xspace}
\newcommand{\methodname}{D4C\xspace}
\newcommand{\correctfix}{180}
\newcommand{\plaufix}{210}

\newcommand{\modify}[1]{\textcolor{black}{#1}}

\def\BibTeX{{\rm B\kern-.05em{\sc i\kern-.025em b}\kern-.08em
    T\kern-.1667em\lower.7ex\hbox{E}\kern-.125emX}}

\begin{document}

\title{\modify{Aligning the Objective of LLM-based Program Repair}}

\author{\IEEEauthorblockN{
    Junjielong Xu\IEEEauthorrefmark{2}, 
    Ying Fu\IEEEauthorrefmark{4}, 
    Shin Hwei Tan\IEEEauthorrefmark{6}, 
    Pinjia He\IEEEauthorrefmark{2}\IEEEauthorrefmark{1}
    \thanks{\IEEEauthorrefmark{1}Pinjia He is the corresponding author.}\vspace{2ex}
    }
\IEEEauthorblockA{
    \IEEEauthorrefmark{2}The Chinese University of Hong Kong, Shenzhen, China\quad \IEEEauthorrefmark{4}Chongqing University, China\quad\IEEEauthorrefmark{6}Concordia University, Canada}
\IEEEauthorblockA{
    junjielongxu@link.cuhk.edu.cn\quad fuying@cqu.edu.cn\quad shinhwei.tan@concordia.ca\quad hepinjia@cuhk.edu.cn}
}

\maketitle
\begin{abstract}
Large language models (LLMs) have achieved decent results on automated program repair (APR).
However, the next token prediction training objective of decoder-only LLMs (\eg GPT-4) is misaligned with the masked span prediction objective of current infilling-style methods, which impedes LLMs from fully leveraging pre-trained knowledge for program repair.
In addition, while some LLMs can locate and repair bugs in certain functions using the related artifacts (\eg test cases), existing methods still depend on statement-level fault localization methods to provide a list of buggy hunks for repair.
This restriction hinders LLMs from exploring potential patches beyond the given locations.

In this paper, we investigate a new approach to adapt LLMs to program repair.
Our core insight is that LLM's APR capability can be greatly improved by simply aligning the output to their training objective and allowing them to refine the whole program without first identifying faulty statements.
Based on this insight, we designed \methodname, a straightforward prompting framework for APR.
\methodname can repair \correctfix{} bugs correctly in Defects4J, with each patch being sampled only 10 times. 
This surpasses the SOTA APR methods with perfect fault localization by 10\% and reduces the patch sampling number by 90\%.
Our findings reveal that (1) objective alignment is crucial for fully exploiting LLM's pre-trained capability, and (2) replacing the traditional localize-buggy-hunks-then-repair workflow with direct debugging is more effective for LLM-based APR methods.
Thus, we believe this paper introduces a new mindset for harnessing LLMs in APR.
\end{abstract}


\input{Sections/001_Introduction.tex}

\input{Sections/002_Motivation.tex}
\input{Sections/003_Approach.tex}

\input{Sections/004_Evaluation.tex}

\input{Sections/005_Conclusion.tex}

\balance
\bibliographystyle{IEEEtran}
\bibliography{ref}

\end{document}

%% file: Sections/001_Introduction.tex
\section{Introduction}\label{sec:intro}

Program repair is a critical part of the software cycle. 
To fix bugs in software systems, developers often need to dedicate substantial time (exceeding 35\% of regular development time) for manual program repair~\cite{humandebugcost}.
To reduce such human effort, researchers have started exploring Automatic Program Repair (APR) methods~\cite{aprsurvey,learningaprsurvey}. 
Based on how the patches are generated, APR methods can be categorized into heuristic-based~\cite{genprog,capgen}, constraint-based~\cite{spr,angelix}, template-based~\cite{tbar,sketchfix}, and learning-based methods~\cite{sequencer,recoder}.
Traditional methods often rely on pre-defined patterns for solving specific bug types (\eg NullPointerException).
Learning-based methods typically use large-scale, high-quality bug-fix pair data to train neural machine translation (NMT) models~\cite{seq2seq}, which transform a buggy program to a fixed version~\cite{learningaprsurvey}.
Recently, as Large Language Models (LLM)~\cite{gpt3} have shown strong code understanding abilities~\cite{cao2025should,wan2024mrweb,wang2024asclepius,Wang2025ASO}, many researchers have begun exploring LLM-based APR methods~\cite{alpharepair,ring}.
With broad pre-training across diverse general corpus, LLMs can achieve impressive repair performance through prompt engineering~\cite{can,li2025reasoning} or minimal data fine-tuning~\cite{impact}.

\begin{figure*}[tbp]
\centering
\includegraphics[scale=0.48]{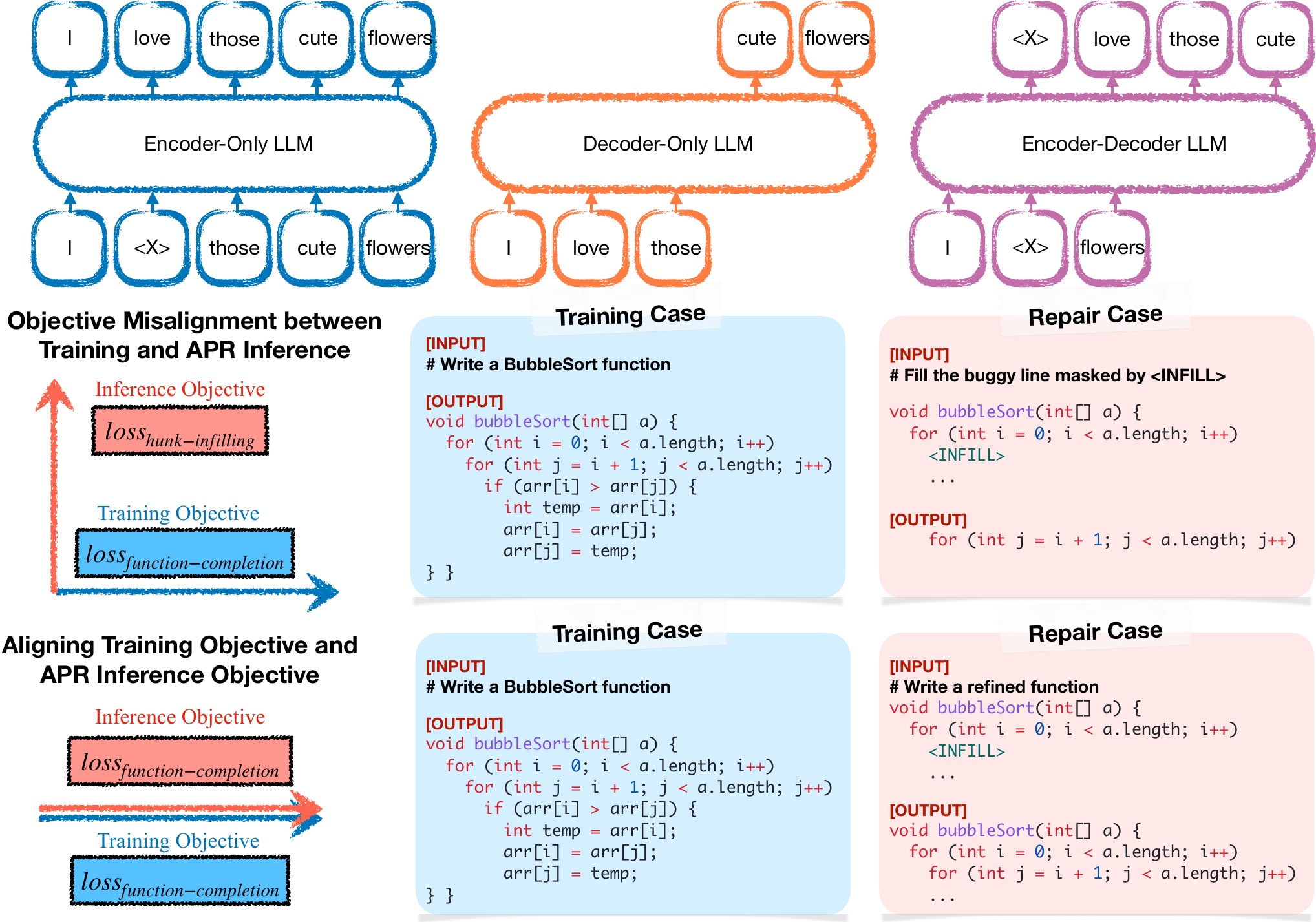}
\caption{
\textbf{First row}: LLM sturctures and their training objectives.
\textbf{Second row}: The training and inference objective is misaligned when using decoder-only LLMs for infilling-style APR.
\textbf{Third row}: An intuitive way to align the gap: using LLMs for entire program completion rather than masked span prediction.
}
\label{fig:objectiveexample}
\end{figure*}

However, current application of LLMs is not well-adapted for program repair due to two major reasons.
\textbf{\textit{First}, the inference objective of current approaches is misaligned with LLM's training objective.}
LLMs can be classified into \textit{encoder-only}, \textit{encoder-decoder}, and \textit{decoder-only}.
The first two are typically trained to predict the masked tokens (\ie \textit{infilling}~\cite{bert} or \textit{denoising}~\cite{t5}), while the third is trained to predict the next tokens (\ie \textit{completion}~\cite{gpt3}).
Existing LLM-based methods~\cite{alpharepair, rapgen,fitrepair} generally adopt the \textit{infilling}-style method~\cite{alpharepair} to predict the fixed code at the masked buggy code location during inference.
The early work achieved decent results, as the models used are also trained on infilling task~\cite{alpharepair,rapgen}.
Recently, larger decoder-only LLMs like GPT-4~\cite{gpt4} have shown stronger capability on code tasks~\cite{sparks}, and people attempt to directly employ them for infilling-style APR~\cite{llmaprstudy,chatrepair,conversational,lee2024unified}.
However, while the model parameters being scaled up by hundreds of times, the fixed bugs did not even doubled~\cite{alpharepair,chatrepair}, which contrasts with the clear scalability in other tasks~\cite{sparks}.
We hypothesize that this is due to decoder-only LLMs are trained for completion rather than infilling, resulting in an objective misalignment between training and inference.
Specifically, due to the sparsity of parallel \texttt{<masked, denoised>} code corpus in training, these LLMs are less proficient in patch generation on masked buggy code.
As noted in recent studies~\cite{adaptingqa,adaptingcls}, objective misalignment can yield significant sub-optimal performance.

\textbf{\textit{Second}, the current workflow limits LLM from fully exploiting its pre-trained capability.}
Existing approaches still adhere to the workflow of first using \modify{statement-level} fault localization (\modify{statement-level} FL) tools~\cite{flsurvey} to obtain a ranked list of potential \modify{buggy hunks (\ie contiguous statements to be modified in patches~\cite{hunk})}, and then using APR methods to generate patches by modifying the buggy program at these \modify{hunks}~\cite{aprsurvey,learningaprsurvey}.
However, LLMs already have the capability to fix trivial syntax bugs in their generated code independently by referring to different artifacts (\eg error report from compilers and virtual machines)~\cite{ring,agentcoder,selfdebug}, illustrating the capability of identifying bugs without \modify{given buggy hunks}.
Considering LLM's superior code comprehension ability, and several studies show that current FL tools may fail to provide precise \modify{buggy hunks}~\cite{flstudy,liu2019you}, asking LLMs to generate patches at provided \modify{several hunks} might contrarily hinder them from exploring broader potential patch space beyond the provided ones.
A recent study also revealed that LLMs fail to make good use of the given buggy lines and tend to over-rely on them~\cite{impact}, whereas another study suggested using a more flexible fault localization to enhance APR~\cite{fan2023automated}.
As observed in these studies, the traditional \modify{localize-buggy-hunks-then-repair} workflow may lead to ineffective LLM-based APR.

In this work, we explore a novel way to use LLM for program repair.
\textbf{\textit{We propose that (1) aligning the output from infilling discrete hunks to completing entire functions can better attain the training objective, and (2) allowing LLM to locate and repair buggy hunks with artifacts in a human-like manner can further improve its APR performance.}}
Based on these insights, we implemented \textbf{\textit{\methodname}} (\ie \textbf{D}irect \textbf{D}ebug \textbf{D}rives \textbf{D}ecent \textbf{C}ode), a straightforward APR approach without complex prompting or extra training.
We conducted experiments on a total of 1027 bugs associated with individual functions (\ie single-function bugs) from Defects4J~\cite{defects4j} and DebugBench~\cite{debugbench}.
\methodname repaired \correctfix{} out of the 437 single-function bugs in Defects4J, outperforming the state-of-the-art (SOTA) APR methods with perfect \modify{statement-level} FL~\cite{chatrepair} by 10\%.
Meanwhile, each patch in \methodname was sampled only 10 times, accounting for 10\% used by SOTA methods (which use 100--5000 samples). 
Our results show that by fully aligning the output format and augmenting the input prompt, \methodname can achieve the best APR performance without \modify{given buggy hunks}, additional fine-tuning, or multi-round dialogue.
This paper is not aimed at proposing \methodname as a new APR technique, but rather to introduce a new mindset or paradigm for better harnessing LLMs in APR in the future.

In summary, this paper makes the following contributions.
\begin{description}[leftmargin=*]
        \item[Problem reformulation:] We reformulate infiling-style repair problem as a program refinement problem. Our experiments show that asking LLMs to generate an entire refined function can align with the pre-training objective of decoder-only LLMs, leading to SOTA APR performance. 
        \item[New APR workflow:] Instead of relying on \modify{statement-level FL tools} to obtain the list of \modify{buggy hunks for repair} as in traditional APR workflow, we show that allowing LLMs to locate and repair \modify{buggy hunks} by using diverse types of artifacts (\eg failed tests, error messages, and code comments) in a human-like manner can further improve its APR performance.
        \item[Implementation and evaluation:] We implemented \methodname based on these insights, and the evaluation results on two widely used benchmarks (\ie Defects4J and DebugBench) have demonstrated its effectiveness.
\end{description}

%% file: Sections/002_Motivation.tex
\section{Background and Motivation}\label{sec:background}

\subsection{LLM Architectures and Training Objectives}
\textbf{Background}: A large language model (LLM) is a language model consisting of a neural network with many parameters (typically billions of weights or more) trained on large quantities of unlabelled corpus using self-supervised learning~\cite{llm}.
The LLMs usually adopt the Transformer~\cite{transformer} architecture or one of its sub-structures (\ie \textit{encoder} or \textit{decoder}).
\modify{The encoder usually consists of feed-forward networks with self-attention~\cite{transformer}, while the decoder usually consists of feed-forward networks with cross-attention~\cite{transformer}.}
Thus, LLMs can be categorized into three types: \textit{encoder-only}, \textit{decoder-only}, and \textit{encoder-decoder} LLMs.

\textit{Encoder-only LLMs}, such as BERT~\cite{bert} and its variants like CodeBERT~\cite{codebert}, have a bidirectional transformer encoder structure.
They are typically trained on the masked language modeling objective (\ie MLM), aiming to denoise and reconstruct the masked tokens via understanding the surrounding context (Fig.~\ref{fig:objectiveexample}).
As shown in Eq.~\ref{eq:mlm}, the loss of MLM training objective can be explicitly represented as the log-sum of the conditional probabilities of generating the masked token when all the unmasked tokens are known.
\begin{equation}\label{eq:mlm}
    \mathcal{L}_{MLM} = -\frac{1}{M}\sum_{i\in M} \log P(\hat{t}_{i}=t_{i}| t_{k\notin mask}; \theta)
\end{equation}
where $M$ is the total number of masked tokens.

\textit{Decoder-only LLMs}, including GPT series~\cite{gpt3,gpt4} and LLaMA series~\cite{llama}, have an autoregressive transformer decoder structure.
They are mainly trained on the causal language modeling objective (\ie CLM), aiming to predict and complete next tokens via following the prefix context (Fig.~\ref{fig:objectiveexample}).
As shown in Eq.~\ref{eq:clm}, the loss of CLM training objective can be explicitly represented as the log-sum of the conditional probabilities of generating the next token when all the preceeding tokens are known.
\begin{equation}\label{eq:clm}
    \mathcal{L}_{CLM} = -\frac{1}{N}\sum_{i\in N} \log P(\hat{t}_{i}= t_{i}| t_{1}, t_{2}, ... , t_{i-1}; \theta)
\end{equation}
where N is the total number of the input tokens.

\textit{Encoder-decoder LLMs}, such as T5~\cite{t5} and its variants like CodeT5~\cite{codet5}, have a complete transformer structure.
They use an encoder to embed the input text into vector representations, and a decoder to causally generate new tokens after the input prompt.
Specifically, T5 series LLMs are often pre-trained on the denoising objective, which aims to \modify{generate the reconstructed spans (\ie a sequence of adjacent tokens) from the masked spans} for the masked span in the prompt (Fig.~\ref{fig:objectiveexample}).
As shown in Eq.~\ref{eq:denoising}, the loss of denoising training objective can be explicitly represented as the log-sum of the conditional probabilities of generating the masked token when all input tokens are known.
\begin{equation}\label{eq:denoising}
    \mathcal{L}_{Denoising} = -\frac{1}{M}\sum_{i\in M} \log P(\hat{t}_{i}= t_{i} | t_{k\in input}; \theta)
\end{equation}
where M is the total number of the masked tokens.
While this process is similar to the MLM objective of encoder-only LLMs, it only generates the reconstructed tokens from the masked position, ignoring other parts (\eg ``I'' and ``flowers'' in Encoder-Decoder LLM in Fig.~\ref{fig:objectiveexample}). Similar to decoder-only LLMs, it also has an unfixed length of text generation.

\begin{figure}[tbp]
\centering
\includegraphics[scale=0.48]{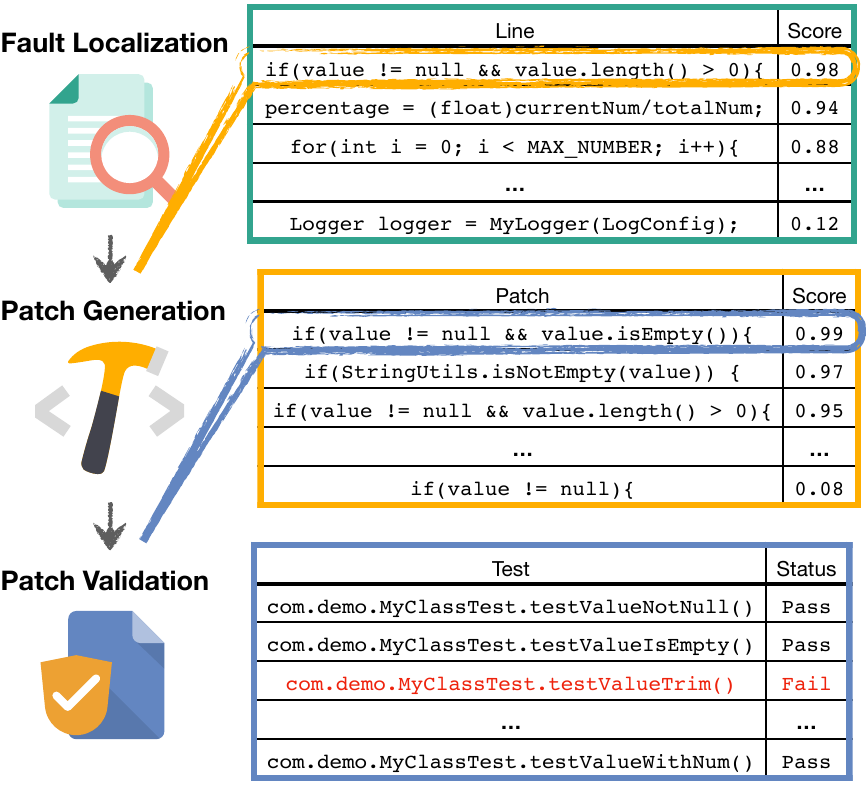}   
\caption{
An example of \modify{automated program repair workflow.}
}
\label{fig:workflowexample}
\end{figure}

\textbf{Motivation}: As introduced, both the encoder-only and encoder-decoder (especially T5 series) LLMs involve the corruption and denoising process during pre-training, which substitutes some tokens in the input text for mask tokens, and reconstruct these tokens from the mask tokens. 
To make effective use of them for APR, Xia~\etal~\cite{alpharepair} proposed the infilling-style APR. 
It aims to replace the entire input buggy hunk with a mask token (\eg \texttt{<INFILL>}) and prompts the model for direct inference, restoring the fixed hunk from the mask. 
The infilling-style APR can be implemented differently across various models. 
For instance, in CodeBERT, it recovers the correct code from masked hunks while leaving the rest of the input tokens the same~\cite{codebert}.
In CodeT5, it only outputs the correct code restored from the masked hunks~\cite{codet5}.

\begin{figure*}[tbp]
\centering
\includegraphics[scale=0.6]{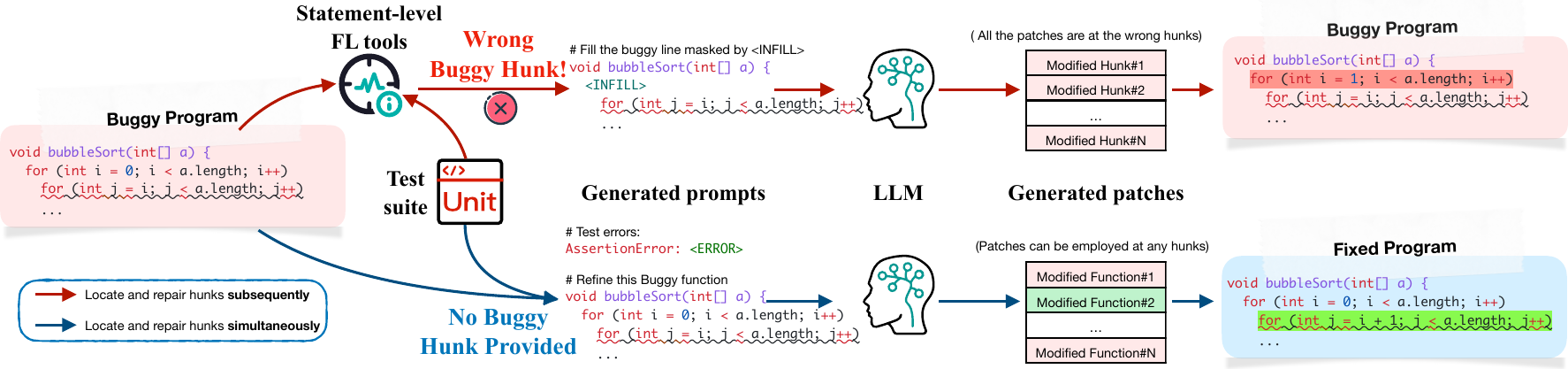}   
\caption{
An example of two different APR paradigms.
\textbf{First row}: Locate and repair \modify{buggy hunks} \textit{subsequently} may cause many invalid attempts for patch generation. 
\textbf{Second row}: Locate and repair \modify{buggy hunks} \textit{simultaneously} can mitigate the cost of patching at specific hunks. (The wavy line is a \modify{buggy hunk})
}
\label{fig:paradigmcompare}
\end{figure*}

Recently, several APR approaches use decoder-only LLMs as they have exhibited a stronger coding capacity.
Influenced by the prior success of infilling, it has been generally assumed that such methods would be useful for decoder-only LLMs.
Thus, they use decoder-only LLMs to generate fixed hunks for the given buggy hunks, as illustrated in Fig.~\ref{fig:workflowexample}.
However, except for very few models (\eg InCoder~\cite{incoder}), decoder-only LLMs barely incorporate denoising objective during pre-training. 
As the \texttt{<masked, denoised>} parallel corpus is rarely available, it becomes challenging for the models to learn the generation of the fixed hunk from the masked tokens without further adaptation after the next token prediction pre-training.
Such an objective misalignment might hamper LLM's performance on specific tasks extensively, such as the text classification task~\cite{adaptingcls} and the QA task~\cite{adaptingqa}, whose inference objectives are misaligned with the training objective of decoder-only LLMs.
Thus, we have the following insight.
\begin{mybox}
\textbf{Insight 1}: Modifying decoder-only LLMs' output from fixed hunks to the entire refined program can better align the inference objective to the training objective, thus significantly enhancing APR performance.
\end{mybox}

This insight was inspired by previous approaches focusing on optimizing model performance on specific tasks by aligning the pre-training objectives of the model~\cite{adaptingcls,adaptingqa}.
We will verify our hypothesis in Sec.~\ref{sec:rq2} via comparing completion perplexity of two output formats (\ie fixed hunks or complete function) on white box LLMs (\ie Mixtral-MoE).

\subsection{APR Techniques and Workflow}

\textbf{Background}: 
Given a buggy program and an artifact (e.g., usually a test suite with at least one failed test), automated program repair (APR) approaches generate a fixed program that fulfills a correctness criteria (e.g., passing all tests).
As illustrated in Fig.~\ref{fig:workflowexample}, \modify{an APR workflow} typically encompasses three steps: (1) fault localization, (2) patch generation, and (3) patch validation. 
Particularly, APR research mainly focuses on the patch generation step and obtains the fault locations using existing FL techniques (\eg statistical fault localization) or uses perfect FL results in evaluation.

Based on their patch generation strategies, APR methods can be categorized into heuristic-based~\cite{genprog,capgen,tan2015relifix,an2018comparing,yuan2018arja,jiang2018shaping,le2016history}, constraint-based~\cite{angelix,nguyen2013semfix,xuan2016nopol}, 
template-based~\cite{kim2013automatic,tbar,sketchfix}, and learning-based~\cite{cure,sequencer,recoder,ye2022selfapr,ye2022neural,jin2023inferfix,ahmed2023better}. 
The non-learning-based approaches are usually restricted by a limited set of program transformations, causing the generation of a large number of invalid patches.
To identify location to apply these transformations, the current strategy is to first generate a ranked list of suspicious \modify{buggy hunks} using \modify{statement-level} FL tools.
The APR tool then sequentially generates patches for each provided \modify{hunks}.
Each of the generated patches is then validated using the given artifacts (\eg test cases).
With the advent of deep learning (DL), learning-based APR tools based on neural machine translation (NMT) have emerged. 
These tools outperform the traditional methods~\cite{learningaprsurvey} via learning code semantics on a large bug-fix parallel training corpus without specially designed patch templates nor search heuristics. 
Moreover, since NMT models generally have a limited parameter size, their training costs and inference efficiency are manageable. 
Therefore, they are well-suited for the traditional APR workflow.

\begin{figure*}[tbp]
\centering
\includegraphics[scale=0.55]{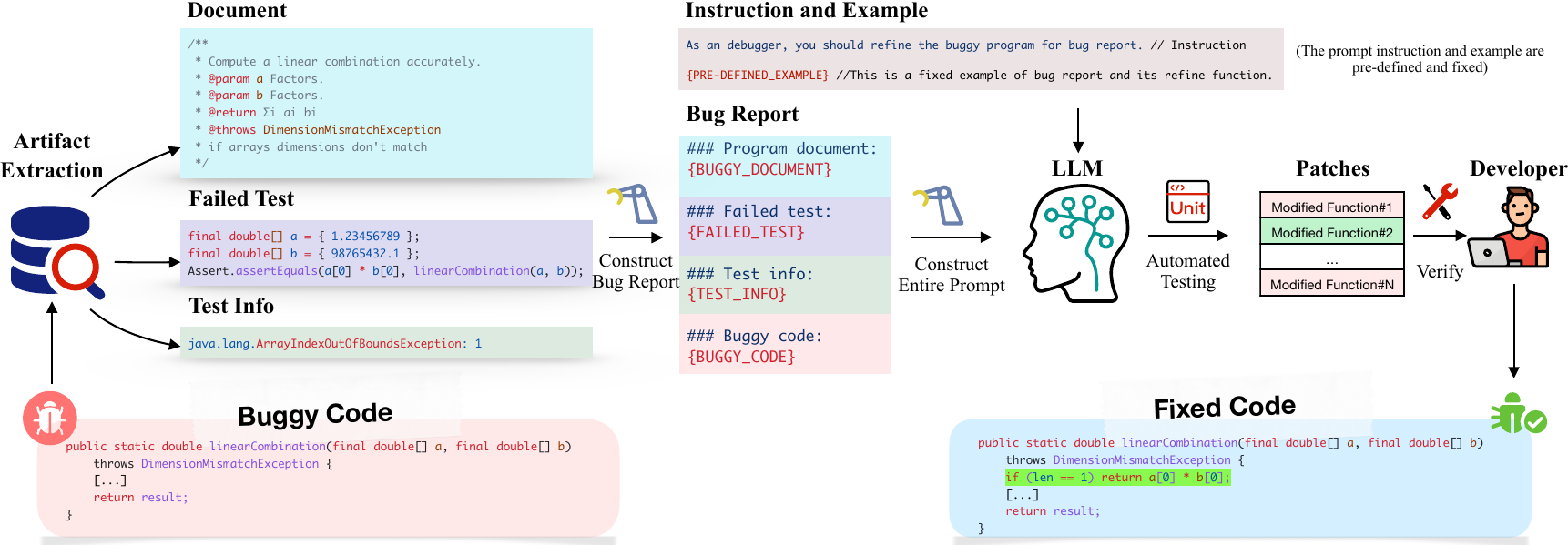}   
\caption{
The workflow of \methodname. It uses the buggy code and its corresponding documents, failed tests, and test info (\eg error message) to construct the prompt for one-shot prompting-based program repair without a specific \modify{buggy hunk} (usually provided by \modify{statement-level} FL tools).
}
\label{fig:d4c}
\end{figure*}

\textbf{Motivation}: In recent years, there is a growing interest in using LLMs (\eg GPT-4~\cite{gpt4}) for APR~\cite{alpharepair,rewardrepair,fan2023automated,jin2023inferfix}. 
Researchers usually follow the current APR workflow, positioning LLM as a new patch generator to replace former NMT models.
However, taking LLM as a simple substitute in the patch generation step in the current workflow is an under-utilization of its pre-trained knowledge, since these models exhibit the capability of locating and fixing bugs in buggy functions independently.
For example, SELF-DEBUGGING~\cite{selfdebug} has been proposed with the concept of allowing LLMs to progressively refine their generated buggy functions to produce bug-free functions by engaging in multiple self-dialogue rounds with artifacts like execution traces, without using statement-level FL to identify buggy hunks.
As shown in Fig.~\ref{fig:paradigmcompare}, since \modify{statement-level} FL tools may not always provide perfect predictions of \modify{buggy hunks}, restricting LLM to fixing the provided \modify{hunks} may lead to time waste in validating patches at incorrect locations.
Moreover, as LLM's inference overhead and time cost is much higher than traditional APR tools, using LLM to generate patches for all given \modify{hunks} may result in huge resource waste.
Thus, we have the insight below:
\begin{mybox}
\textbf{Insight 2}: Prompting LLMs with buggy programs and corresponding artifacts can enable them to locate and repair \modify{buggy hunks} simultaneously without \modify{statement-level} FL, thus further improving APR performance.
\end{mybox}
This insight arises from the APR methodology of programmers who typically identify \modify{buggy hunks} through tests and documents, and subsequently fix bugs based on these test results.
Moreover, existing research revealed that models can produce decent fixes on basic compile bugs when supplemented with compiler errors~\cite{ring}, and their patch generation capability can be further improved with the assistant of the given failed test information~\cite{chatrepair,selfdebug}.
\modify{Thus, we can use a more flexible FL method~\cite{fan2023automated} to identify the buggy segments of the program at a coarse-grained level (\eg using method-level FL~\cite{fluccs} to find the buggy function), and allow the LLM to simultaneously locate and fix the buggy hunks within those segments, as locating the segments containing the buggy hunks is easier and more practical than using statement-level FL to directly find the correct buggy hunks for repair.}
We will verify our hypothesis in Sec.~\ref{sec:rq2} via comparing the number of correct patches of two input formats (\ie w/ or w/o artifacts).

%% file: Sections/003_Approach.tex
\section{\textbf{\methodname}: \textbf{D}irect \textbf{D}ebug \textbf{D}rives \textbf{D}ecent \textbf{C}ode}\label{sec:approach}

Based on these two insights, we developed an APR framework \methodname, \ie direct debug drives decent code. 
As shown in Fig.~\ref{fig:d4c}, when presented with a new buggy program\footnote{In our implementation, due to the LLM's token length limitation, a buggy program refer to a buggy function \modify{that can be provided by method-level FL}.}, \methodname uses the related artifacts, including the documents and failed test information, to construct a bug report.
Then, \methodname will use this bug report to instruct the model to generate a refined version of the buggy program.
\modify{To align LLM's inference objective with its pre-trained completion objective, we adopt a one-shot prompting strategy. This involves prefixing a fixed example of buggy program and its refined version before the target buggy program in the prompt, enabling LLM to infer their input-output relations and finally generate a refined version of target program in expected format.}
\methodname does not require any additional fine-tuning.
It only changes the output format to align with the pre-training objectives, and uses test information and documents (\eg Javadoc comments) in the input prompt to enable repair without the prior knowledge of the correct \modify{buggy hunks}.
We implemented our approach on GPT-4 series models and Mixtral-MoE models.
Next, we will introduce our problem definition (Sec.~\ref{sec:pd}) and the design details of \methodname, including model selection (Sec.~\ref{sec:model}), artifact extraction (Sec.~\ref{sec:input}), prompt construction (Sec.~\ref{sec:output}), patch generation (Sec.~\ref{sec:patchgeneration}) and validation (Sec.~\ref{sec:patchvalidation}).

\subsection{Problem Definition}\label{sec:pd}

We model the APR task as a \textit{completion} task that aims at generating a complete refined program based on a buggy program and associated artifacts.
This is different from previous LLM-based APR approaches, which treated the APR task as an \textit{infilling} or \textit{cloze} task~\cite{alpharepair}.
\modify{Notably, the artifacts are essential for \methodname{} to identify the location to modify, while infilling APR methods rely on statement-level fault localization to pinpoint buggy hunks. Although the artifacts are also used in some approaches~\cite{chatrepair} to enhance repair, they are not necessarily required by all infilling methods.}
Specifically, the optimization objective of \methodname can be formally written as:
\begin{equation}\label{eq:completion}
    {\arg\max}_\theta\ \ P(t_\text{Fixed Program}|t_\text{Buggy Program},t_\text{Artifacts};\theta)
\end{equation}
and the objective of existing infilling-style APR is:
\begin{equation}\label{eq:infilling}
    {\arg\max}_\theta\ \ P(t_\text{Fixed Hunk}|t_\text{Buggy Program},\modify{t_\text{Artifacts}};\theta)
\end{equation}
where $t$ means the tokens, $\theta$ is the trainable parameters of LLMs. 
The reason for the difference between \methodname and the previous methods in task modeling is that \methodname is designed to exploit the pre-training capability of decoder-only LLMs trained on CLM objective, whereas infilling-style APR is designed to exploit the pre-training capability of encoder-only LLMs trained on MLM objective or encoder-decoder LLM trained on denoising objective.

\subsection{Model Selection}\label{sec:model}

The LLM itself is the most vital part of LLM-based applications.
Since the goal of proposing \methodname is to illustrate an adaptive way to use current advanced decoder-only LLMs for APR, the backbone of \methodname should be a decoder-only LLM which has been pre-trained on a substantial code corpus via next token prediction objective.
Moreover, \methodname does not aim at teaching extra knowledge to LLMs for APR.
Instead, its APR ability primarily comes from mining the pre-trained knowledge via aligning the model's response with its training corpus, \eg a complete function. 
Thus, to enable \methodname exhibit the state-of-the-art APR performance, we choose the most advanced LLMs, GPT-4, to serve as the backbone in our implementation.
However, since GPT-4 is a black-box model whose inference loss of CLM objective (\ie \textit{perplexity}) is unavailable, we have to choose another state-of-the-art white-box model, Mixtral-MoE, as alternative backbone to validate whether generating a whole refined function can better align the training objective. We select Mixtral-MoE because at the time of submission, the strongest model on the HumanEval leaderboard~\cite{leaderboard} with the longest context window was Mixtral-MoE.
We will not dive into their architecture or pre-training details in this paper, but it is worth noting that the backbone for \methodname can be changed to other LLMs with next token prediction (CLM) training objective.
To best ensure our replicability, we employed fixed remote API checkpoints or fixed local model versions.
For reference, we also provide comparative experiments about using different backbones in Sec.~\ref{sec:rq2}.

\subsection{Artifact Extraction}\label{sec:input}

Before patch generation, \methodname needs to extract related artifacts for subsequent prompting.
As shown in Fig.~\ref{fig:d4c}, \modify{once the buggy function is identified,} \methodname will automated extract (1) documents or comments that describe the general purpose of the function and its input-output data types, (2) the inputs and expected outputs of the failed test cases, and (3) error messages from executing those failed test cases.
Our underlying motivation for providing these artifacts to the LLM is that LLM has gradually exhibited human-like analysis and reasoning abilities recently~\cite{agentcoder}.
Instead of treating LLM as a tool as in the traditional mindset of training NMT models from scratch, we should treat it as an intelligent agent.
From this perspective, we contemplate how humans debug, and present this procedure to LLM for execution.
Specifically, in the human program repair process, we need to first intuitively understand the function's purpose so we refer to related code documents or function-level comments. 
Then, we wonder what input can reproduce the bug.
Thus, we will check the relevant failed test cases and error messages.
Finally, in the absence of communication or confirmation from other developers (\ie in a \textit{rubber duck debugging}~\cite{rubberduck} cases), we can only use these artifacts to simultaneously locate and repair the bug.
Therefore, we extract artifacts and construct bug report for LLM to repair.
 
Notably, the implementations of the artifacts extraction procedure may differ due to various data characteristics.
For instance, although we default to using function-level comments as documents, if certain functions do not have this artifact, \methodname can alternatively exploit related \texttt{README} documents. 
Moreover, if some artifacts are unavailable in some scenarios (\eg the test cases of the online judgment are unknown), \methodname will replace these in the table with a placeholder statement, like ``\textit{This program does not possess any known test cases.}''

\subsection{Prompt Construction}\label{sec:output}

\begin{figure}[tbp]
\centering
\includegraphics[scale=0.7]{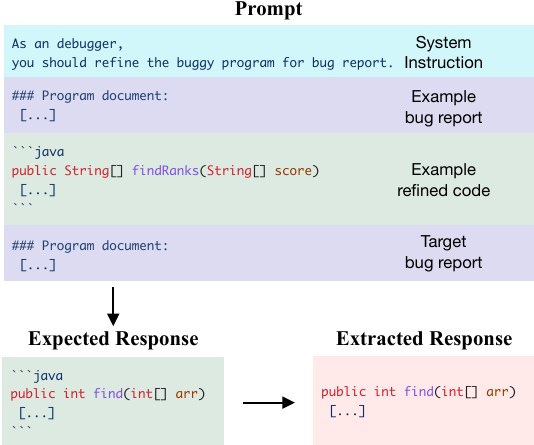}   
\caption{
The prompt structure of \methodname. The details of the code are omitted.
\modify{The example pair is fixed, which is used to constrain LLM's response format.}
}
\label{fig:prompt}
\end{figure}

After the extraction of artifacts, \methodname assembles these artifacts into a comprehensive bug report. 
As shown in Fig.~\ref{fig:d4c}, the report template is concise, with only the most basic notations of the different components in the bug report (\eg program and comment), without providing any complex instructions or requirements.
This allows LLM to focus more on the bug report, thus accomplishing our goal of inducing and exploiting LLM's pre-trained knowledge and capabilities for program repair, rather than ``teaching'' the model to repair.
Following this, \methodname integrates the bug report into a complete prompt, which is carefully designed to include a concise role-play system instruction (\eg \textit{You are an AI debugger ...}) and a \modify{fixed} example consisting of a handcrafted bug report and its refined program. 
Our underlying goal is to stimulate LLM's in-context learning (ICL) ability~\cite{gpt3}, \ie \modify{analogizing the report-refine pair in the prompt to further understand the instruction.} 
This one-shot ICL also allows us to restrict the output format of the LLM to a \textit{refined function} rather than other common output formats related to APR (\eg unit diff patch~\cite{diff}).
Moreover, such restricted output format assists us in automatically extracting the corrected functions from LLM response for automated patch verification as illustrated in Fig.~\ref{fig:prompt}.
We only use one fixed example to make LLMs follow the output format since our goal is using \methodname to show the effectiveness of our insight, rather than developing a powerful method with complex prompting strategy like RAG~\cite{unilog,divlog}.

Notably, the implementation of prompts may vary across different models.
Specifically, in the local text completion LLMs, the prompt can be treated as a single text input.
In the remote chat models that support multi-turn dialogues, each module of the prompt represents an individual dialogue message in the conversation.
For instance, in OpenAI models, instructions are encapsulated within the structure body of system messages, while the input and output of examples are encapsulated within the structure body of user messages and assistant messages respectively. 
In open-sourced Mixtral, instructions are marked via special tokens \texttt{[INST]} and \texttt{[/INST]}, while the remaining content is split into different messages using \texttt{<\textbackslash s>} as separators.

\subsection{Patch Generation}\label{sec:patchgeneration}
Unlike traditional APR tools that generate patches specifically for locations that need modification, the patch generation step of \methodname refines the entire program in the inference stage.
This is primarily because the core goal of \methodname is to explore the extent of APR enhancement by aligning the inference objective to decoder-only LLM's pre-training objective.
Furthermore, we do not introduce an additional patch ranking method, which is widely adopted by current APR approaches.
This is because \methodname only needs to generate at most \textit{10} patches during decoding (\ie beam search), which is significantly less than the sampling number of existing methods (\ie over \textit{hundreds} of sampling).
Thus, \methodname can subsequently verify all 10 patches \modify{without involving significant waste of effort on incorrect patches (within \textit{5 minutes}), rather than suffering from a} time-consuming process of verifying hundreds of patches \modify{(up to \textit{5 hours}~\cite{alpharepair,chatrepair,rapgen,repilot,fitrepair})}.
Additionally, although \methodname can rank the patches by comparing their naturalness (quantified by perplexity or entropy)~\cite{naturalness,localness}, a prior study indicates that it is inaccurate to identify the correctness of the program using naturalness~\cite{unnaturalness}. 
Therefore, we do not use any re-ranking strategy to save the computational cost incurred by the perplexity calculation.
Following previous LLM-based work that focus on method-level generation~\cite{du2024evaluating}, we assume that the \textit{buggy function} is provided where \methodname can select lines within the function to modify.
In a real-world scenario, the buggy function can be provided by existing method-level FL tools~\cite{fluccs} via analyzing the execution results of failed tests.

\subsection{Patch Validation}\label{sec:patchvalidation}

After generating the candidate patches, \methodname extracts the refined program and uses them to replace the buggy program in the original source code.
Then, \methodname runs the corresponding test suite to find patches that compile successfully and pass all tests.
However, due to the potential incomplete test coverage problem of benchmarks, the APR methods may often produce patches that pass all tests without exactly meeting the expected functionality of the developer~\cite{anti}, thus do not truly fix the bug.
These patches are known as \textit{plausible patches}.
Ideally, developer validation is helpful, but the bugs in Defects4J are historically fixed and outdated~\cite{defects4j}, making developer feedback impractical. 
To ensure rigor experiment, we follow existing APR papers~\cite{alpharepair,repilot,rapgen,fitrepair,chatrepair} by adding a manual validation for each plausible patch after automated evaluation to identify the correct patches that are the same or semantically equivalent to the human-written patch.

%% file: Sections/004_Evaluation.tex
\section{Evaluation}\label{sec:validation}

This section aims to answer the research questions below:

\begin{itemize}
    \item \textbf{RQ1: How does D4C compare against LLM-based APR methods?} 
    We compared \methodname to state-of-the-art APR methods \modify{with and without} perfect FL. We aim to investigate the improvement from aligning output and enhancing input as mentioned in Sec.~\ref{sec:intro}.
    \item \textbf{RQ2: How effective are the two insights that drive the design of D4C?} 
    We aim to show that generating a refined function better aligns LLM's training objective, and using the artifacts to locate and repair bugs without \modify{given buggy hunks} further exploits LLM's APR ability.
    \item \textbf{RQ3: How do different components and parameter affects D4C?}
    We conduct an ablation study and a sensitivity analysis on \methodname. We aim to quantify the contribution of each prompt component and characterize the performance of \methodname in different parameter settings.
\end{itemize}

\subsection{Experiment Setup}

\subsubsection{Environment and Implementation}

We use a black-box LLM (\ie \texttt{gpt-4-0613}) and a white-box LLM (\ie \texttt{mixtral-8x7b-instruct-v0.1}) via OpenAI APIs~\cite{openaiapi} and 8xA100 NVIDIA GPU server for program repair.
We use Python 3.9 to implement the inference and evaluation scripts in a local machine with Ubuntu 20.04.5 LTS.
To enhance LLM's instruction following ability, we manually crafted a \modify{fixed} bug-fix example.
This example is employed in each experiment to construct 1-shot prompting \modify{for all bugs}.
To adapt to different experiments, we modified its input-output format according to the corresponding experiment requirement.
\modify{To control a moderated randomness of text generation, we follow the previous work and set the randomness hyper-parameter, \texttt{temperature}, as \texttt{1.0}. This is also the default value of OpenAI~\cite{openaiapi}.}
\modify{To facilitate patch validation, we set a timeout threshold of 1 minutes for each patch.}
Furthermore, we adopt a much smaller sampling number of \texttt{10}, as the inference budgets (fee, time, etc.) for LLMs are much larger than traditional methods.
Unless otherwise specified, all experiments adhered to this setting.

\begin{table}[htbp]
\caption{Statistics of the benchmarks: The number of single-funciton bugs and the threat of data leakage}
\setlength\tabcolsep{3.5pt}
\centering
\label{table:statistic}
\scalebox{1}{
\begin{tabular}{cccc}
    \toprule
    Benchmark & Bug number  & Data leakage & Language\\
    \midrule 
    \modify{Defects4J} & 437 & Yes & Java\\
    DebugBench & 200/194/196 & No & C++/Java/Python3\\
    \bottomrule
\end{tabular}
}
\end{table}

\subsubsection{Datasets}\label{sec:datasets}

Our experiments are conducted on 1027 logic single-function bugs from DebugBench (590) and bugs from Defects4J (437).
Specifically, DebugBench~\cite{debugbench} is a new debug benchmark designed to counter data leakage (by implanting bugs into source data with GPT-4), which contains a total number of 4,253 bugs from Java, C++, and Python3 from LeetCode.
In general, fixing logic bug is more difficult than fixing other types of bugs. 
For example, syntax bugs can be easily detected by the compiler or interpreter without testing, and their fixing methods are usually provided in the raised exceptions. 
Thus, our paper aligns with the practice of existing APR methods that only focus on fixing logic bugs.
Since DebugBench is not exposed to the threat of data leakage, we use it for our insight validation (RQ2).
However, due to the complexity to reproduce current baselines on this new benchmark, we do not use it for the comparison between \methodname and baselines.
Defects4J~\cite{defects4j} is a widely used APR benchmark with 835 real-world bugs from 17 open-source repositories.
Following previous work~\cite{llmaprstudy,chatrepair,conversational}, we separate the single-function bugs of Defects4J to v1.2 (203 bugs) and v2.0 (234 bugs).
Since it contains data from previous versions of open-source projects where most code models have been trained on, it faces the threat of data leakage.
However, this threat is believed to be less severe in APR compared to other domains~\cite{impact} due to the sparsity of bug-fix parallel corpus in the training data, meaning that model can only learn the individual buggy or fixed version of the program, not their pairs.
Despite this, the potential threat of data leakage could still undermine the validity of hypothesis verification.
Hence, Defects4J is mainly used as a reference for comparing \methodname with existing baselines.
The statistic of the benchmarks are shown in Table~\ref{table:statistic}.

\subsubsection{Metrics}\label{sec:metric}

Following the previous work, we use the number of \textit{correct patch} to evaluate the APR effectiveness.
Specifically, if a patch can pass all unit tests, then it will considered as a plausible patch.
If this plausible patch truly resolves the bug, rather than overfit to pass the unit test only (\eg via referring the provided failed tests in the prompt), then it will be confirmed as a correct patch.
In our evaluation, we adhere to prior studies~\cite{alpharepair,chatrepair,repilot} and manually identify whether a plausible patch truly resolves the bug.
Notably, the bugs from DebugBench are collected from LeetCode, and the LeetCode programs that can pass all the unseen test suites from the website are deemed as correct programs.
In our experiment, we can only access the \textit{provided test examples} of each LeetCode problem in our bug report, rather than the \textit{failed unseen test} from LeetCode OJ.
Thus, we also determine the patches that can pass all the LeetCode tests as correct.
To distinguish from manually checked correct patches, we refer to patches can pass the LeetCode validation as \textit{verified patches}.

\subsubsection{Baselines}\label{sec:baseline}

In RQ1, we selected several state-of-the-art LLM-based APR methods as baselines, including AlphaRepair~\cite{alpharepair}, FitRepair~\cite{fitrepair}, ChatRepair~\cite{chatrepair}, RAP-Gen~\cite{rapgen}, and Repilot~\cite{repilot}. 
We did not compare \methodname with learning-based APR methods trained from scratch since our goal is to illustrate the effectiveness of \methodname among all the LLM-based approaches.
All these baselines employ the infilling objective.
In RQ2, we chose different input and output format for comparison to show the superiority of aligning output format and providing artifacts in the input.
We also chose prompts that add or delete different information (\eg function comments) as baselines to investigate the contributions of different information to APR.

\subsection{RQ1: Comparison against LLM-based APR}\label{sec:rq1}

We compare \methodname with five state-of-the-art LLM-based APR approaches on Defects4J, the most commonly used benchmark for APR.
Since some of the baselines are not open-sourced, we cannot re-run them to reproduce the result.
Thus, we reuse their Defects4J results reported in the original paper.
Although using Defects4J may have the the risk of exposing its data to LLM's training corpus, we still use it for evaluation because (1) existing work~\cite{impact} has shown that data leakage is less of a concern for APR compared to other code-related tasks as the training corpus often contains at most the individual versions of the program (\ie learning buggy or fixed version only), and LLMs can hardly learn the APR tasks without bug-fix pair corpus during training; and (2) the baselines have also been evaluated on Defects4J (hence, they are subject to the same data leakage risk).
Following prior work, we retain this setting to ensure fairness of the comparison.
The comparison result is shown in Table~\ref{table:main} \modify{and Table~\ref{table:fl}}.

\begin{table}[thbp]
\caption{
The number of correct patches generated by different approaches on Defects4J \modify{v1.2 \& v2.0}.
}
\label{table:main}
\centering
\scalebox{1}{
\begin{tabular}{cccccc}
    \toprule
    \textbf{Method} & \textbf{Model} & \textbf{v1.2} & \textbf{v2.0} & \textbf{Sum} & \textbf{Sampled}  \\
    \midrule
    AlphaRepair & CodeBERT & 52 & 34 & 86 & 5000 \\ 
    Repilot & InCoder & 66 & 50 & 116 & 5000 \\
    RAP-Gen & CodeT5 & 72 & 53 & 125 & 100 \\
    FitRepair & CodeT5 & 89 & 44 & 133 & 5000 \\ 
    ChatRepair & GPT-4 & \textbf{114} & 48 & 162 & 100-200\\ 
    \midrule 
    \methodname & GPT-4 & 84 & \textbf{96} & \textbf{180} & \textbf{10} \\
    
    \bottomrule
\end{tabular}
}
\end{table}

\begin{table}[thbp]
\caption{
\modify{The number of correct patches generated in Perfect FL and Statistical FL on Defects4J v1.2}.
}
\label{table:fl}
\centering
\scalebox{1}{
\begin{tabular}{cccccc}
    \toprule
    \textbf{Method} & \textbf{Model} & \textbf{Perf.} & \textbf{Stat.} & \textbf{Drop} & \textbf{Sampled}  \\
    \midrule
    AlphaRepair & CodeBERT & 52 & 36 & 30.8\% & 5000 \\ 
    RAP-Gen & CodeT5 & 72 & 48 & 33.3\% & 100 \\
    \midrule 
    \methodname & GPT-4 & 84 & \textbf{80} & \textbf{4.8\%} & \textbf{10} \\
    
    \bottomrule
\end{tabular}
}
\end{table}

\modify{In Table~\ref{table:main},} we calculated the number of correct patches on Defects4J (v1.2, v2.0, and their sum) and patches sampled per bug (notated as ``Sampled") \modify{in the perfect localization settings}.
\modify{Specifically, for baselines, the perfect \textit{buggy hunks} within the buggy function is provided, while for \methodname, only the \textit{buggy function} is provided.}
Notably, \methodname largely outperforms all the existing state-of-the-art methods provided with \textit{\modify{perfect buggy hunks}} for infilling-style APR.
\methodname (GPT-4) can generate almost 10\% more correct patches than ChatRepair, the latest APR methods which uses GPT-4 for multi-run dialogue with provided \modify{buggy hunks}.
Moreover, \methodname only needs to sample 10 times for each bug in patch generation, which is 90\% fewer than that required by the most efficient baseline (sampling at least 100 patches).
The fewer generated patches also illustrate the better efficiency of \methodname and less waste of cost in patch validation.
\modify{
However, using only perfect FL settings does not demonstrate the flexibility and practicality of \methodname, which does not rely on given buggy hunks. 
To illustrate its effectiveness in real-world scenarios, we further conducted experiments of \methodname and baselines using existing statistical FL tools.
}

\modify{
In Table~\ref{table:fl}, we report the number of correct patches on Defects4J under perfect FL (Perf.) and statistical FL (Stat.) settings. 
The baselines use statement-level FL tools (\eg Tarantula~\cite{flsurvey}) to identify buggy hunks, whereas \methodname employs a method-level FL tool (\ie FLUCCS~\cite{fluccs}) to locate buggy functions.
We compare \methodname only with AlphaRepair and RAP-Gen on Defects4J v1.2 for two reasons: (1) these are the only methods that reported the evaluation results with non-perfect FL setting on Defects4J v1.2 in the paper, and (2) FLUCCS’s code supports only Defects4J v1.2. 
In our experiments, \methodname generates only 10 patches for the most suspicious faulty function (Top@1 method) given by FLUCCS.
Notably, \methodname repairs only 4 fewer bugs compared to the perfect FL settings, with a minimal drop of 4.8\% in repair success. 
This is significantly lower than the infilling-style APR baselines, which experienced at least a 30\% drop. 
These results suggest that \methodname is less affected by inaccuracies in FL since locating buggy functions is easier than locating buggy hunks.
}

\modify{Overall, the results in Table~\ref{table:main} and Table~\ref{table:fl} suggest that \methodname is effective and efficient.}
However, due to the potential of data leakage, we cannot directly verify our insight of \methodname via comparing current approaches.
Thus, we conduct more in-depth experiments on DebugBench in Sec.~\ref{sec:rq2} and Sec.~\ref{sec:rq3}.

\subsection{RQ2: Effectiveness of the two insights}\label{sec:rq2}

RQ2 aims to validate the effectiveness of the two insights introduced in Sec.~\ref{sec:background}, and how the two insights contribute to the overall performance improvement in \methodname.
We design two additional questions for the insights: 
\modify{
(IQ1) \textit{Does allowing LLM to generate a complete refined function better align its training objectives and result in better performance?} 
(IQ2) \textit{Does providing the artifacts enable LLM to locate and repair buggy hunks itself and achieves a better performance?}
}
Specifically, we evaluate the two insights on DebugBench since there is no threat of data leakage. Thus, we use \emph{verified patches} (\ie patches verified by LeetCode unseen tests) in DebugBench for the evaluation. 

\begin{figure}[tbp]
\centering
\includegraphics[scale=0.6]{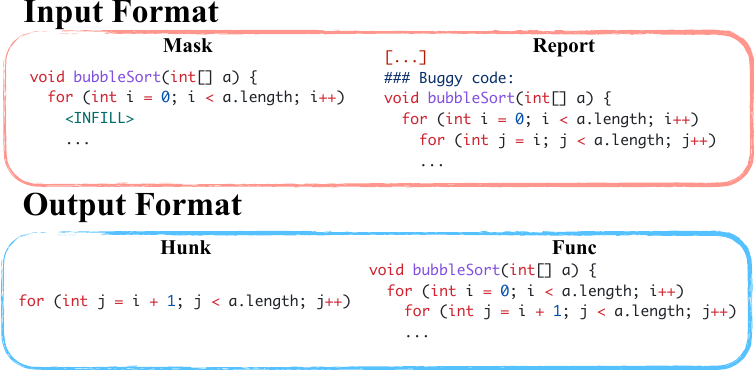}   
\caption{
Examples of input and output format used in RQ2.
}
\label{fig:io}
\end{figure}

\subsubsection{\textbf{Validation of Insight 1}}

To answer the IQ1 question, we calculate the average \emph{\textbf{perplexity}}, \modify{the inference loss of CLM objective (Eq.~\ref{eq:clm}),} for various output format on the white-box model (\ie Mixtral-MoE).
\modify{Measuring perplexity is the most straightforward approach to evaluate whether the generated content aligns with the distribution of training corpus, as LLMs are trained to minimize the CLM objective~\cite{gpt3,llama}.}
\modify{Thus, a lower perplexity value indicate the model is more confident at predicting a given sequence.}
Specifically, we adopt two output formats, including the fixed hunk patches that is used by infilling-style APR (denoted as \emph{Hunk}) and the full-function output used by \methodname (\emph{Func}).
Fig.~\ref{fig:io} shows the examples of these output format.
In our implementation, we also carefully modified the system instructions to adapt to each output setting.
To control variables, we also calculate the perplexity of generated output (O), and the entire input\&output (IO).
Thus, we can compare the perplexity among different output format to validate our insight, ignoring the difference between each input format.

\begin{table}[tbp]
\caption{
Comparison of different settings of \methodname on DebugBench. We measure perplexity and verified patches to validate insights.
}
\label{table:iopplgbj}
\centering
\setlength\tabcolsep{4pt}
\scalebox{1}{
\begin{tabular}{c|cc|ccc|ccc}
    \toprule
    \multirow{2}*{\textbf{Format}} & \multicolumn{2}{c}{\textbf{Perplexity}} & \multicolumn{3}{|c}{\textbf{Mixtral (\#Verified)}} & \multicolumn{3}{|c}{\textbf{GPT-4 (\#Verified)}}\\
    & O & IO & C++ & Java & Python & C++ & Java & Python \\
    \midrule 
    Mask-Hunk  & 8.59 & 2.68 & 66 & 39 & 63 & 140 & 139 & 129 \\ 
    Mask-Func  & 3.01 & \textbf{1.58} & 99 & 91 & 76 & 160 & 153 & 147 \\ 
    \midrule
    Report-Hunk  & 8.50 & 2.64 & 72 & 64 & 67 & 144 & 140 & 133 \\
    Report-Func  & \textbf{1.39} & 1.79 & \textbf{118} & \textbf{125} & \textbf{104} & \textbf{177} & \textbf{163} & \textbf{161} \\ 
    \bottomrule 
\end{tabular}
}
\end{table}

Table~\ref{table:iopplgbj} shows the results.
We observe that when the output is an entire function, the output perplexity is much lower than using the fixed hunks as the output, regardless of the input.
This directly validates that generating a complete function is the better way to align the training objective.
To further validate that it is the objective alignment that helps \methodname achieve a better APR performance, we evaluate the verified patches generated in each case.
The result illustrates that using complete function as output can result in generation of more verified patches with lower perplexity than using discrete hunks.
Moreover, when only generating the fixed hunks as the output, the output perplexity is much higher than its perplexity calculated with the whole input and output sequence.
This indicates that completing discrete hunks only is misaligned with LLM's training objective.
Overall, these findings illustrate the effectiveness of our first insight.

\subsubsection{\textbf{Validation of Insight 2}}

To answer the IQ2 question, we evaluate the patch generation ability on DebugBench using two different input formats, \ie the bug report (notated as Report) and the buggy program where the buggy hunks are masked (Mask).
To control variables, we also conducted experiments using the two previously established output formats as shown in Fig.~\ref{fig:io}.
For the generated patches by \textit{Report-Hunk}, as we do not provide a specific location to modify and cannot precisely determine where the generated patch should be applied, we instruct Mixtral-MoE to help us to automatically apply 
the generated patches to the original buggy program before patch validation.
Table~\ref{table:iopplgbj} shows the number of generated verified patches on both Mixtral-MoE and GPT-4. 
The results indicate that \textit{Report-Func} is the most effective among all evaluated groups, including \textit{Mask-Hunk}, which is the input-output format of infilling-style APR.
This phenomenon aligns with prior evaluations on Defects4J (RQ1). 
It is worthwhile to note in the experiment in Sec.~\ref{sec:rq1}, the effectiveness of \methodname stems not only from the bug report design but also the objective alignment of the output. 
However, even when we use the same output format in this experiment, the performance of using bug report as input still outperforms that of using program where the buggy hunks are masked. 
This suggests that bug report helps not only in \modify{locating the buggy hunks}, but also give more information for patch generation. 
Thus, this phenomenon validates our second insight.

\subsection{RQ3: Ablation study and sensitive analysis}
\label{sec:rq3}

\begin{table}[tbp]
\caption{
The number of patches generated for our ablation study and sensitive analysis. We measure plausible patches for Defects4J and patches verified via unseen tests for DebugBench.
}
\label{table:ablation}
\setlength\tabcolsep{4.0pt}
\centering
\scalebox{1}{
\begin{tabular}{l|crc|cccc}
    \toprule
    \multirow{2}*{\textbf{Benchmark}} & \multicolumn{3}{|c|}{\textbf{Defects4J (\#Plausible)}} & \multicolumn{4}{|c}{\textbf{DebugBench (\#Verified)}}\\
     & v1.2 & v2.0 & Sum & C++ & Java & Python & Sum \\
    \midrule
    w/o Document & 93 & 99 & 192 & 175 & 162 & 158 & 495 \\
    w/o Test & 80 & 89 & 169 & 173 & 159 & 159 & 491 \\
    w/o Message & 81 & 93 & 174 & 169 & 162 & 161 & 492 \\
    - Mask & 53 & 73 & 126 & 160 & 153 & 147 & 460 \\
    \modify{- Pure} & \modify{47} & \modify{62} & \modify{109} & \modify{143} & \modify{140} & \modify{131} & \modify{414} \\
    \midrule
    \midrule
    Samp=1 & 60 & 73 & 133 & 163 & 151 & 145 & 459 \\
    Samp=3 & 91 & 88 & 179 & 177 & 163 & 161 & 501 \\
    Samp=10 (default) & \textbf{96} & \textbf{114} & \textbf{210} & \textbf{181} & \textbf{171} & \textbf{169} & \textbf{521} \\
    \midrule
    Temp=0.0 & 61 & 75 & 136 & 165 & 154 & 146 & 465 \\
    Temp=1.0 (default)  & \textbf{96} & \textbf{114} & \textbf{210} & \textbf{181} & \textbf{171} & \textbf{169} & \textbf{521} \\
    \bottomrule
\end{tabular}
}
\end{table}

In this section, we aim to quantify the contribution of each component, and investigate the sensitivity of \methodname to the sampling number and temperature.
We do not additionally validate the sensitivity on different prompts since we have carefully investigated the impact of different prompt formats in RQ2.
We use GPT-4 as the backbone, set the default temperature to 1.0, and use the sampling number of 10 by default.
As the manual validation of patch correctness on Defects4J is time-consuming, we use the plausible patches in this section. 
We also use verified patches to evaluate the repair performance on DebugBench.
We design the baselines below to check the effectiveness of each component:

\noindent\textbf{w/o Document:} \methodname{} without including the program documents or function comments in the prompt.

\noindent\textbf{w/o Test:} \methodname{} without including inputs and expected outputs of the failed test cases in the prompt.

\noindent\textbf{w/o Message:} \methodname{} without including error message from the failed tests in the prompt.

\noindent\textbf{- Mask:} \methodname{} using masks to identify bugs without including any artifacts in the prompt (\textit{Mask-Func}).

\noindent\textbf{- Pure:} \methodname{} without including any guidance like mask or artifacts to identify bugs in the prompt, \ie an aligned LLM.

Table~\ref{table:ablation} shows the results. 
Overall, the results show that each component is important in guiding \methodname{} in generating plausible/verified patches across the two benchmarks. 
We also observe that the ``- Mask" baseline are the least effective \modify{among all baselines with bug location guidance} because \methodname cannot use any information from the artifact to fix the defect. 
Meanwhile, among the first three baselines where we remove one information from the prompt, we notice that program documents (w/o Document) are the least effective as \methodname is able to generate the greatest number of plausible/verified patches without using the document. 
This is expected as the error message and test are more closely related to the defect than the comments that  merely describe general code features.
\modify{
Without guidance on buggy locations, pure GPT-4 can still fix bugs by following the aligned one-shot example pair to generate a refined function. 
While its effectiveness on Defects4J might be due to potential data leakage during pre-training, DebugBench is not affected by this issue.
Despite this, GPT-4 with the completion objective still fixes more bugs (414 vs 408) on DebugBench than GPT-4 with the infilling objective and masked hunk (``Mask-Hunk" in Table~\ref{table:iopplgbj}). 
This suggests LLMs have learned to write bug-free code during pre-training. 
By aligning the inference task with the pre-trained code completion task, we can effectively leverage their pre-trained knowledge to produce high-quality code.
}

In terms of sample size, we notice that a sample size of 10 is the optimal setting to allow \methodname to find the correct patches. 
We found that increasing the sampling number from 1 to 3 significantly increased the number of plausible and verified patches, but further increasing the sampling number to 10 resulted in a smaller improvement. 
Since \methodname can already achieve satisfactory results when setting the sampling number to 10, we use it as the default sample number for \methodname.
We also observed that setting the temperature to 0 results in fewer plausible and verified patches.
Thus, we adopt a multi-sampling for \methodname rather than greedy decoding (temp=0).

%% file: Sections/005_Conclusion.tex
\section{Discussion}\label{sec:discussion}

\subsection{Threats to Validity}

\noindent\textbf{\modify{External Threats.}} 
One external threat comes from the potential data leakage problem (\ie using \modify{Defects4J} in RQ1, which may have been incorporated in LLM pre-training). We have discussed this threat in RQ1 (\ie data leakage is not significant on APR~\cite{impact}, and the baselines have also been evaluated on Defects4J). To eliminate this threat in our insight validation, we selected DebugBench, a latest, leakage-free benchmark, for evaluation. This not only keeps a fair comparison in RQ1, but also ensures the credibility of our conclusions in RQ2 and RQ3.
As the effectiveness of the model may vary in different settings, our results may not generalize beyond the studied settings and other programming languages beyond the supported ones in Defects4J and DebugBench. We mitigate this threat by reusing configurations in prior work, and widely used benchmarks.

\noindent\textbf{\modify{Internal Threats.}} 
An internal threat lies in the incomplete test coverage problem~\cite{anti} of Defects4J, which do not guarantee the correctness of test-passing patches. To mitigate this threat, we follow the previous work~\cite{alpharepair,chatrepair,fitrepair,repilot} to manually check the correctness of plausible patches.
To mitigate the potential bias in the manual analysis, two authors independently confirm the patch correctness. Any patches with disagreement were presented to the third author for review. However, since we could not submit the patch to the developers of the corresponding repository for confirmation, this threat could not be completely eliminated.
Therefore, we decide to release our experimental results for public verification.

\begin{figure}[tbp]
\centering
\includegraphics[scale=0.5]{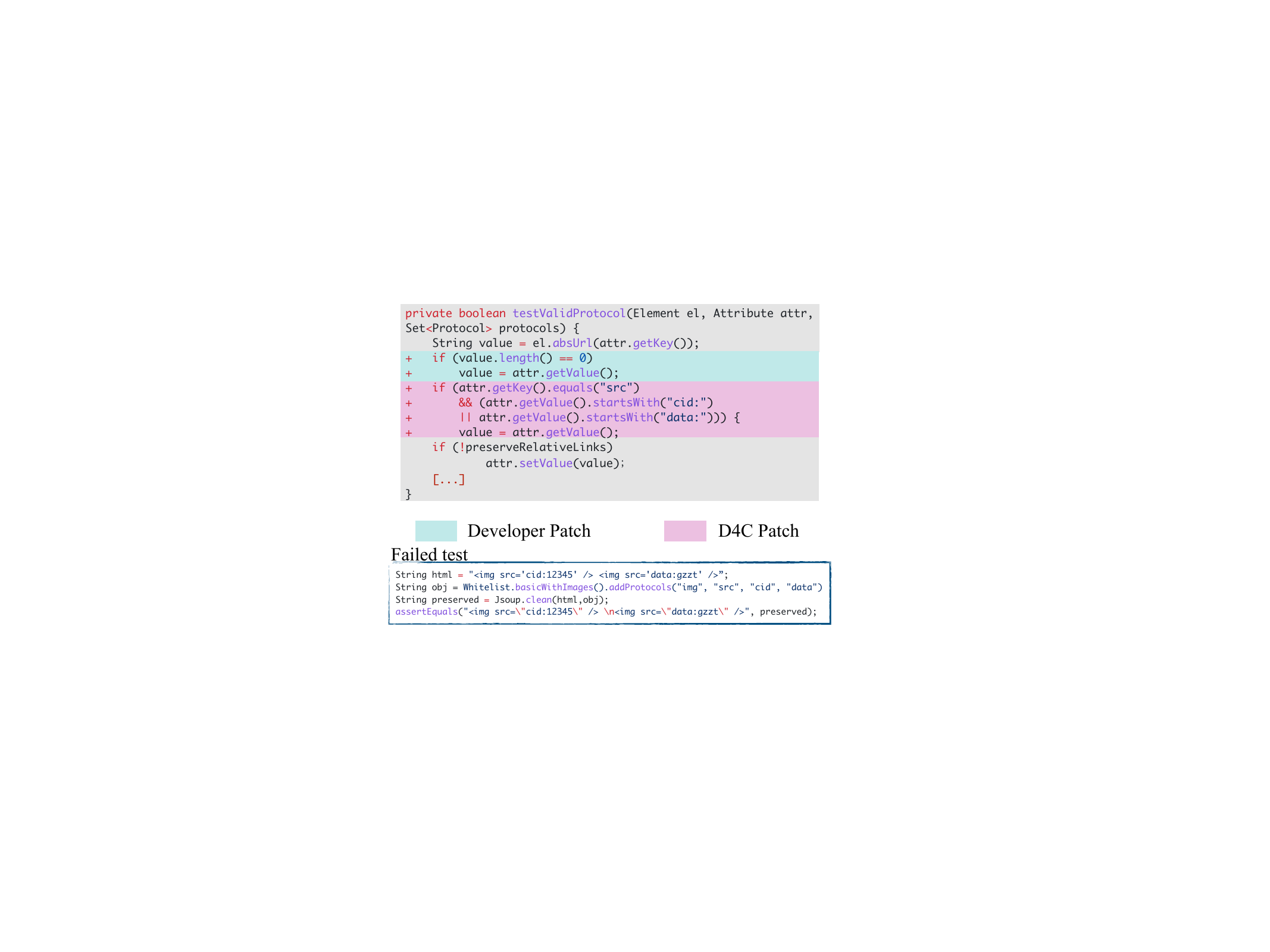}   
\caption{
A plausible patch generated at developer patch locations (Jsoup-19)
}
\label{fig:incorrect}
\end{figure}
\subsection{Qualitative Analysis of Plausible Patches}
\begin{figure}[tbp]
\centering
\includegraphics[scale=0.5]{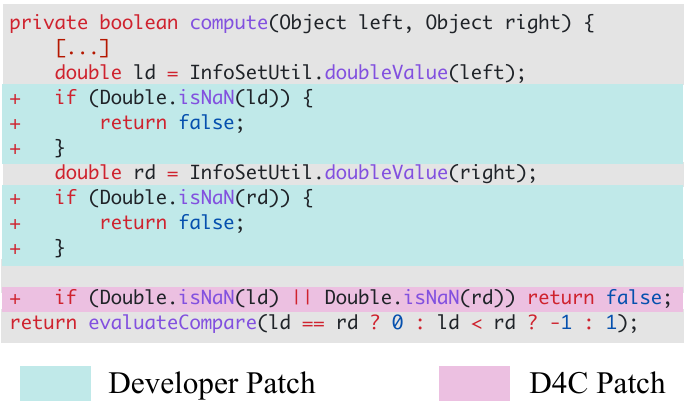}   
\caption{
A correct patch generated at non-developer patch locations (JxPath-8)
}
\label{fig:qua}
\end{figure}

Table~\ref{table:ablation} shows that
\methodname generates a total of \plaufix{} plausible patches but 30 are incorrect with respect to the human patches. 
Fig.~\ref{fig:incorrect} shows one example of the incorrect patch.
Since the failed test contains the \texttt{"src"} tag as input, LLM can naturally generate patches that only pass the test containing \texttt{"src"} without considering patch robustness. This scenario is similar to the overfitting problem in APR~\cite{smith2015cure}.
To check for the overfitting problem, 
two of the authors independently evaluate the patch correctness, followed by a third person to review and confirm their disagreement to ensure the accuracy of the validation.
In our manual validation, we also found that when \methodname can generate five or more plausible patches for a buggy program in 10 samplings, most of these patches are correct. 
This observation aligns with previous work (\ie responses generated multiple times tend to be more reliable~\cite{selfconsistency}).
DebugBench does not suffer from the overfitting problem because the given test examples are not used in its unseen test suite when checking for patch correctness.

As we allow \methodname to debug \modify{without given buggy hunks}, we also observe that some correct patches have been applied at different locations compared to the developer patches.
Fig.~\ref{fig:qua} shows an example where the developer patch modifies two if-conditions so it is a 2-hunk bug in the dataset.
However, \methodname only needs to insert one line, which combines the two conditions and results in a correct patch that is semantically equivalent to the developer patch.
Furthermore, if we ask GPT-4 to generate patches at these two hunks, GPT-4 can only write two \texttt{Exception} statements, and fails to pass the unit tests.
This example shows that allowing LLMs to generate patches without restricting the location may provide more flexibility, allowing generation of correct patches.

\subsection{Token Cost and Price of Patch Generation}\label{sec:price}

The cost of patch generation is a vital criteria to evaluate \methodname's practicality.
For Defects4J, the average input prompt length is 1,386.84 tokens overall, with 1,325.68 for resolved bugs and 1,369.72 for unresolved bugs. The average output completion length is 314.39 tokens, with 285.98 for resolved bugs and 325.80 for unresolved bugs.
Given OpenAI's pricing of \$0.01 per 1,000 input tokens and \$0.03 per 1,000 output tokens~\cite{openaiapi}, the average cost per patch is \$0.023 (\$0.021 for resolved bugs and \$0.024 for unresolved bugs). 
Generating 10 patches per bug results in a total cost of \$0.23 for the entire dataset.
Notably, \methodname fixed 180 bugs in Defects4J, with the correct patch being the 4.29th out of 10 patches generated per bug on average (std.=1.74).
Thus, if we stop patch generation once a correct patch is found, the average cost per bug will be reduced to \$0.18.


\subsection{Wasted Effort of Patch Validation}

The wasted effort of patch validation is another criteria to evaluate \methodname's practicality.
Due to weak test oracles,  not all test-passing (plausible) patches are correct~\cite{qi2015analysis}, and may have anti-patterns~\cite{anti}. 
Thus, there is a labor cost of manually validating the correctness of plausible patches beyond the token cost of patch generation.
Specifically, \methodname generates an average of 5.80 plausible patches per bug (std.=3.28), with every 2.06th plausible patch being correct (std.=1.49).
Compared to infilling APR methods which often require users to manually verify over 400 plausible patches before finding a correct patch with patch ranking (and over 600 without ranking)~\cite{alpharepair}, D4C is significantly more user-friendly.
Even in the theoretical worst case, users only need to validate up to 10 patches per bug, which is significantly fewer than with existing methods.
As discussed in Sec.~\ref{sec:patchgeneration}, this efficiency explains why \methodname does not require the additional re-ranking mechanisms to facilitate patch validation.
Moreover, there is also a time cost for running unit tests during automated patch validation.
However, \methodname requires running tests only up to 10 minutes per bug, with a timeout of 1min per patch.
This is much fewer than a search time limit of up to 5 hours used in previous work~\cite{alpharepair,chatrepair}. 
This illustrates \methodname's efficiency.

\footnotetext{\textbf{Data Availability:} Our source code and experimental results are publicly available at \url{https://github.com/CUHK-Shenzhen-SE/D4C}}

\section{Conclusion}
This paper presents a new approach to adapt LLMs to automated program repair.
Our key insight is that the effectiveness of LLM-based APR can be enhanced by (1) aligning the output to their training objective and (2) allowing them to refine the whole program without \modify{given buggy hunks}.
Based on this insight, we designed \methodname, an LLM-based prompting framework for APR.
Our evaluation shows that \methodname
outperforms the SOTA APR methods with perfect FL by 10\%. Our findings call for adopting a
new paradigm and debugging workflow for future LLM-based APR approaches.

\section{Acknowledgment}
This paper was supported by the Guangdong Basic and Applied Basic Research Foundation (No. 2024A1515010145) and the Shenzhen Science and Technology Program (No. ZDSYS20230626091302006)